\documentclass[a4paper,10pt]{article}
\usepackage{graphicx,wrapfig}
\usepackage{fullpage}
\usepackage{amsmath}
\usepackage[font=small, margin=1cm]{caption}
\usepackage[T1]{fontenc}
\usepackage[utopia]{mathdesign}
\usepackage{color}
\usepackage{siunitx}
\DeclareSIUnit\nano{n}
\DeclareSIUnit\Tesla{T}

\begin{document}

\begin{center}
{\Large{}The HIBEAM Experiment}
\end{center}

\small{\textbf{Alexander Burgman $^{1,}$* ~for the HIBEAM/NNBAR Collaboration}}

$^{1}$ \quad Department of Physics, Stockholm University, 106 91 Stockholm, Sweden

* \quad Correspondence: alexander.burgman@fysik.su.se

\abstract{The violation of baryon number is an essential ingredient for baryogenesis --- the preferential creation of matter over antimatter --- needed to account for the observed baryon asymmetry in the Universe. However, such a process has yet to be experimentally observed.\\
The HIBEAM/NNBAR program is a proposed two-stage experiment at the European Spallation Source to search for baryon number violation. The program will include high-sensitivity searches for processes that violate baryon number by one or two units: free neutron–antineutron oscillation via mixing, neutron-antineutron oscillation via regeneration from a sterile neutron state and neutron disappearance; the effective process of neutron regeneration is also possible. The program can be used to discover and characterize mixing in the neutron, antineutron and sterile neutron sectors. The experiment addresses topical open questions such as the origins of baryogenesis and the nature of dark matter, and is sensitive to scales of new physics substantially in excess of those available at colliders. A goal of the program is to open a discovery window to neutron conversion probabilities (sensitivities) by up to three orders of magnitude compared with previous searches, which is a rare opportunity. A conceptual design report for NNBAR has recently been published.
}

% Keywords
~

%%%%%%%%%%%%%%%%%%%%%%%%%%%%%%%%%%%%%%%%%%

\section{Introduction}

There are several outstanding questions of modern physics that are not explained by the Standard Model of particle physics, including the observed matter/antimatter asymmetry, i.e. that there is an overabundance of matter compared to antimatter in the Universe. In order for the matter/antimatter asymmetry to arise, a set of conditions, specified by Sakharov, must be met~\cite{sakharov_conditions}: (1) the accidental conservation of baryon number, $B$, in the Standard Model must be violated, (2) the charge, $C$-, and charge-parity, $CP$-, symmetries must be violated, and (3) these interactions must occur outside of thermal equilibrium.

The first Sakharov condition, the baryon number violation, $BNV$, can arise with or without the corresponding lepton number, $L$, violation, $LNV$, and one may give rise to the other. Mechanisms for this may involve sphaleron processes, grand unification models, supersymmetry, hidden sector processes, etc., and may be probed via e.g.\ neutrinoless double beta decay searches ($\Delta B=0$, $\Delta L\neq0$, $\Delta \left[B-L\right]\neq0$), proton decay searches ($\Delta B\neq0$, $\Delta L\neq0$, $\Delta \left[B-L\right]=0$), or searches for neutron conversion into antineutrons or hidden sector neutrons ($\Delta B\neq0$, $\Delta L=0$, $\Delta \left[B-L\right]\neq0$)~\cite{Calibbi:2016ukt}. The latter was last probed in the 1990s~\cite{Baldo-Ceolin:1994hzw}, and is the subject of the proposed HIBEAM/NNBAR program at the European Spallation Source~\cite{Addazi:2020nlz}.

The European Spallation Source (ESS) ERIC is a research infrastructure under construction in Lund, Sweden, that will provide the world’s highest intensity source of neutrons for studies in, for example, material science, life science or fundamental physics~\cite{Abele:2022iml}. The ESS is jointly hosted by Sweden and Denmark, with 13 member states from around Europe, and plans on hosting 22 instrumented beamlines (of which 15 are currently decided).

The ESS neutrons will be produced through the spallation process on a rotating tungsten target wheel using a \SI{2}{\GeV} pulsed proton beam, with \SI{14}{\Hz} repetition and \SI{3}{\ms} pulses, and an initial (final) power of \SI{2}{} (\SI{5}{}) \SI{}{\MW}. The cold neutrons are then guided by moderators to the experimental beamlines, which are accessed through several beam ports around the circumference of the target area. Three standard beam ports have been merged to form one larger beam port (dubbed the large beam port) to allow maximum neutron flow, which can be used in fundamental physics studies with the planned NNBAR program\footnote{The NNBAR studies have been conducted as a part of the Horizon2020-funded HighNESS program.}~\cite{NNBAR_CDR}.

\begin{figure}[h]
\centering
\includegraphics[width=10.5 cm]{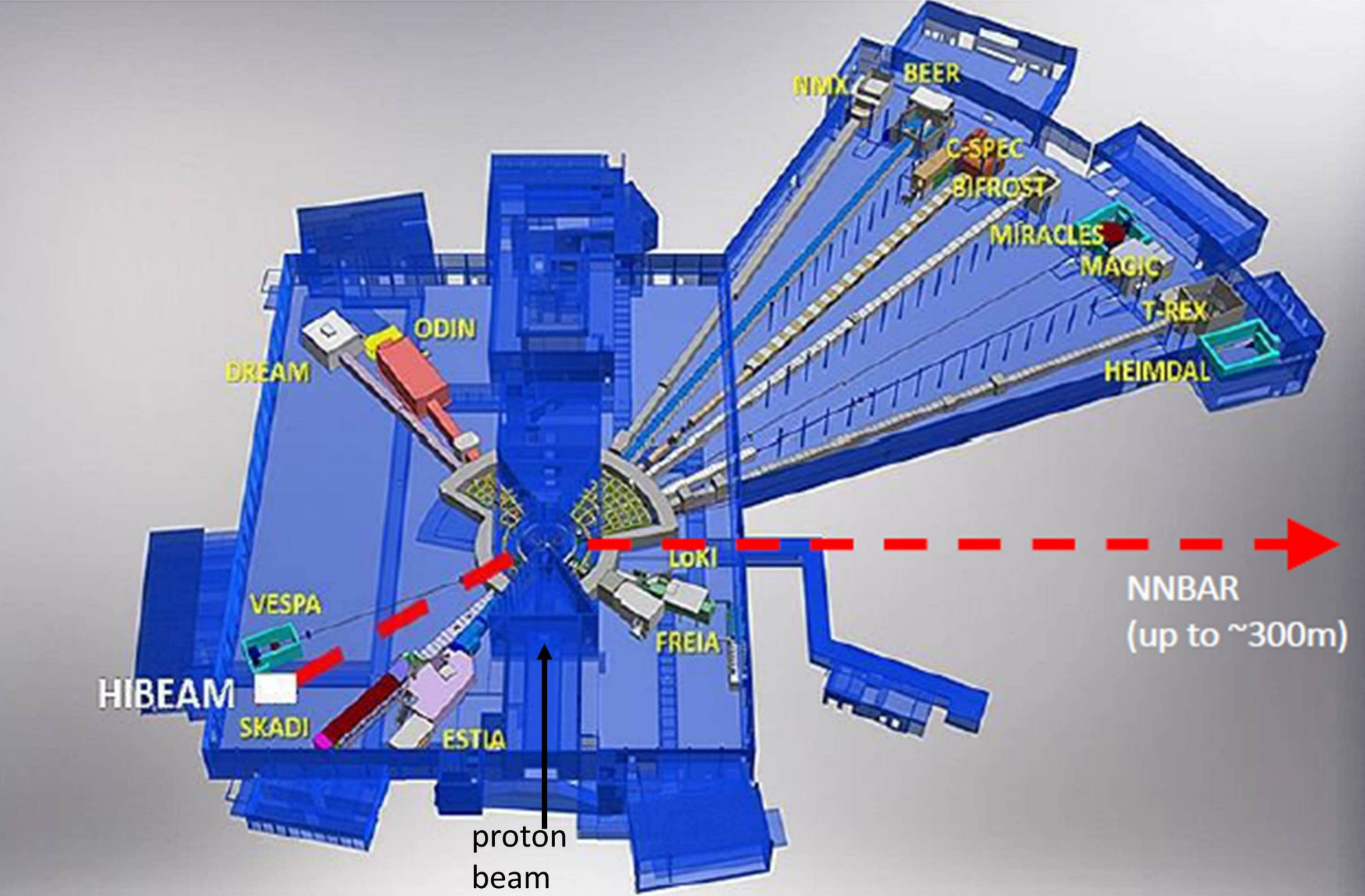}
\caption{Illustration of the ESS experimental hall~\cite{Addazi:2020nlz}.
HIBEAM is proposed to be located to the bottom left of the figure, and NNBAR towards the right. 
\label{fig:ess_hall}}
\end{figure}   
%\unskip

%%%%%%%%%%%%%%%%%%%%%%%%%%%%%%%%%%%%%%%%%%
\section{The HIBEAM/NNBAR Program}

The HIBEAM/NNBAR program~\cite{Addazi:2020nlz,Santoro:2023izd,NNBAR_CDR,Backman:2022szk} will search for neutrons $n$ produced at the ESS that convert into antineutrons $\bar{n}$ or sterile/mirror neutrons $n’$ in a two-stage program, first with the HIBEAM experiment, increasing the discovery potential for these processes by up to one order of magnitude compared to the last search, followed by the NNBAR experiment, which will increase this discovery potential with more than a factor of 1000~\cite{Gonzalez:2024dba}. NNBAR has been detailed in a recently released conceptual design report~\cite{NNBAR_CDR}.

Figure~\ref{fig:ess_hall} shows the layout of the ESS experimental hall, including the proposed placements for the HIBEAM and NNBAR beamlines. NNBAR is placed at the large beam port.

\subsection{The NNBAR Experiment}

\begin{figure}[t]
\centering
\includegraphics[width=13.5 cm]{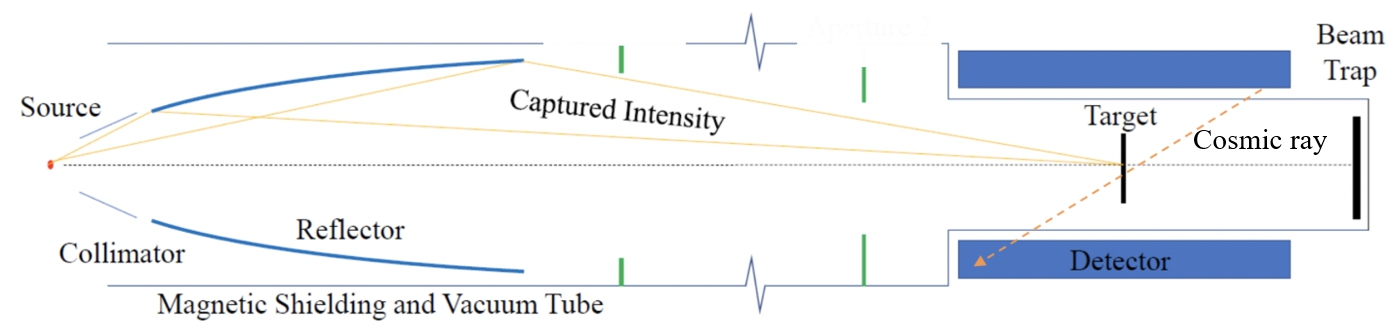}
\caption{Illustration of the NNBAR beamline~\cite{NNBAR_CDR}. 
\label{fig:nnbar_beamline}}
\end{figure}

Figure~\ref{fig:nnbar_beamline} shows the layout of the NNBAR experiment, from the neutron source at the large beam port, through the neutron reflector optics for beam focusing, into the oscillation tunnel, and finally incident on a carbon target foil inside the annihilation detector. The \SI{200}{\m} long oscillation tunnel must be under vacuum and magnetic field free ($<\SI{10}{\nano\Tesla}$) in order to allow the oscillation between neutrons and antineutrons or mirror neutrons. This will be achieved using a passive shield of mu-metal. In order to maximize oscillation time, the neutrons will propagate very slowly, $\lesssim\SI{1000}{\m\per\s}$. When an oscillated neutron reaches the target foil it will annihilate with a neutron in a carbon nucleus into 4--5 pions (charged and neutral), with a center-of-mass energy of $\sqrt{s}=2m_n\approx\SI{1.9}{GeV}$. The pions, spread isotropically, will each carry an energy of $\lesssim\SI{500}{\MeV}$ (peaking around \SI{100}{\MeV}), and be detected with the surrounding detector, consisting of a time-projection chamber, a lead-glass electromagnetic calorimeter, and a scintillator-based hadronic range calorimeter. Detector prototypes have been constructed and will be tested~\cite{Dunne:2021arq}. 

Multiple studies have been conducted to demonstrate the ability to reconstruct the pion multiplicity, mass and the system’s invariant mass, as well as to separate signal pions from background protons and cosmic rays, using the NNBAR detector~\cite{Barrow:2021deh,billy_detector}. The neutron-antineutron (or  mirror neutron) oscillation would be a very rare process, which requires an excellent separation between signal and background events. This is achieved through highly effective event selection, which yields a rejection of background events by more than a factor of $\SI{E9}{}$ while maintaining a signal efficiency of $\sim\SI{70}{\percent}$~\cite{NNBAR_CDR}.

%\unskip

\begin{figure}[b!]
\centering
\includegraphics[width=10.5 cm]{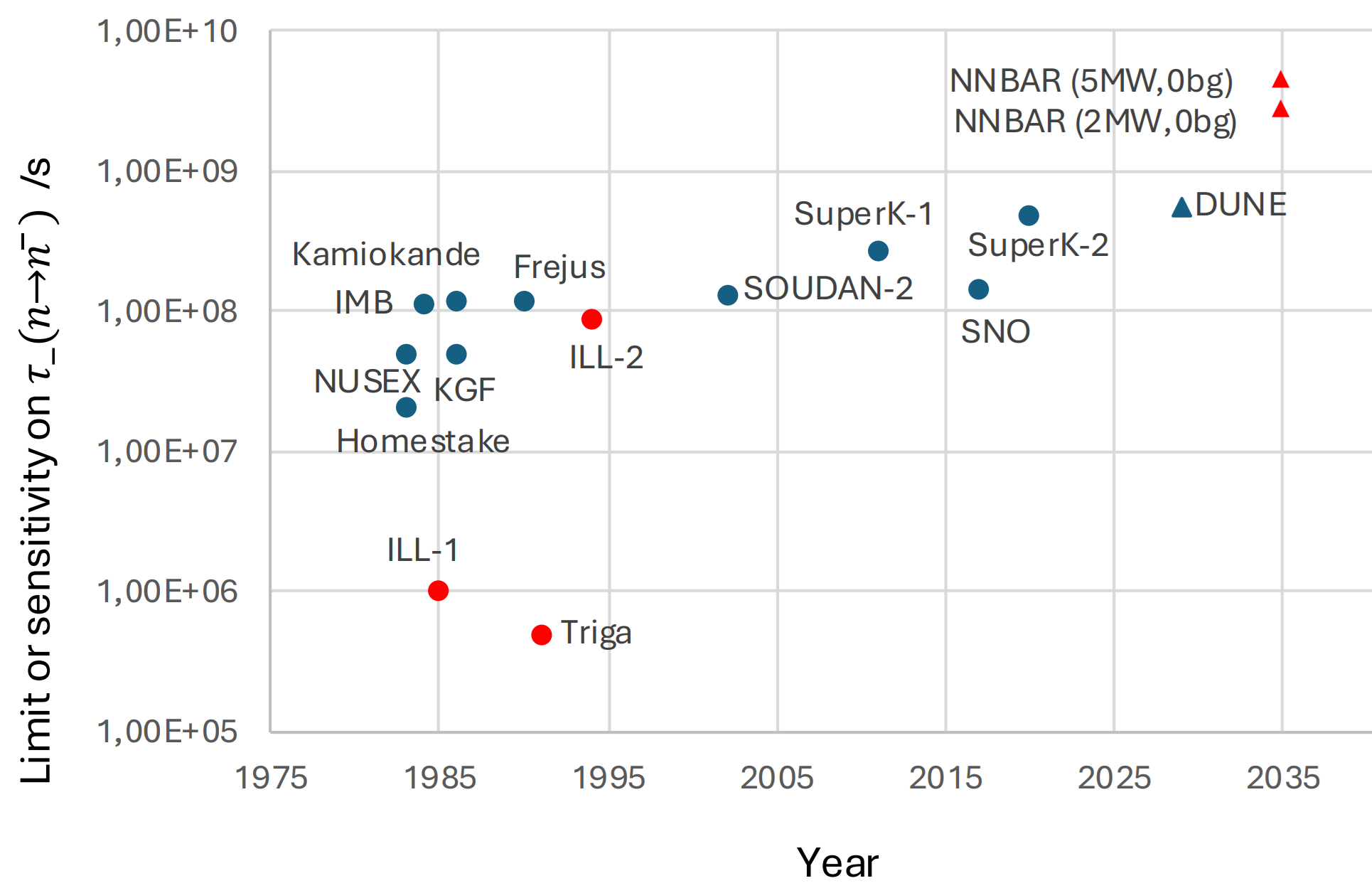}
\caption{The NNBAR sensitivity to $n\rightarrow\bar{n}$ oscillation compared to previous experiments (circles) and future experiments (triangles).
Indirect searches (blue) are shown along with direct searches (red)~\cite{NNBAR_CDR}.
\label{fig:nnbar_sensitivity}}
\end{figure}

The resulting estimated sensitivity is shown in Figure~\ref{fig:nnbar_sensitivity} compared to previous experiments. This surpasses the most recent upper limit produced using free neutrons by 1,5 orders-of-magnitude, and the most recent upper limit using bound neutrons (highly model dependent) by one order-of-magnitude~\cite{NNBAR_CDR}.

%\unskip

\subsection{The HIBEAM Experiment}

The HIBEAM experiment is the precursor to NNBAR. It will make use of a magnetically shielded $\sim\SI{50}{\m}$ beamline at the ESS, and can be operated in four distinct search modes: (a) direct neutron-antineutron oscillation, $n\rightarrow\bar{n}$, (b) neutron-mirror neutron oscillation, $n\rightarrow n’$, (c) neutron to mirror neutron to neutron oscillation, $n\rightarrow n’\rightarrow n$, and (d)  neutron to mirror neutron to antineutron oscillation, $n\rightarrow n’\rightarrow \bar{n}$. Search modes (c) and (d) make use of a mechanical beam stop in the middle of the beamline to stop the large flow of neutrons while allowing the non-interacting mirror neutrons to pass. Modes (a) and (d) operate neutron annihilation detectors using a carbon foil (similar to NNBAR), either with a bespoke detector or the WASA crystal calorimeter, while modes (b) and (c) make use of a neutron counting detector. This will result in a discovery potential improvement of a factor up to 10 compared with previous results, both for antineutron and mirror neutron oscillation~\cite{Santoro:2023izd}.

\begin{figure}[h]
\centering
\includegraphics[width=10.5 cm]{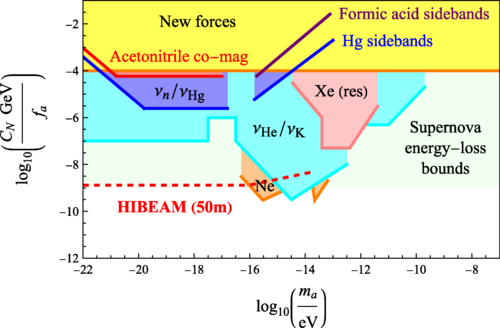}
\caption{Sensitivity of the HIBEAM experiment to direct detection of ambient axions, shown as a function of the axion mass~\cite{Fierlinger:2024aik}.
\label{fig:hibeam_axions}}
\end{figure}

Additionally, HIBEAM will be sensitive to axions over a broad range of possible axion masses~\cite{Fierlinger:2024aik}. The ambient axions would act as a pseudo-magnetic field to the neutrons in the beamline, and therefore change their Larmor frequency (magnetic moment precession), which can be detected through a Ramsey interferometry experiment. This yields a sensitivity to direct axion detection comparable to the indirect supernova energy-loss limit for axion masses from $\SI{E-22}{\eV}$ to $\SI{E-16}{\eV}$, which is an improvement of more than two orders of magnitude over previous direct searches (see Figure~\ref{fig:hibeam_axions}).

%\unskip

%%%%%%%%%%%%%%%%%%%%%%%%%%%%%%%%%%%%%%%%%%
\section{Summary and Outlook}

Discovering $BNV$ in the neutron sector, as neutron oscillation into antineutrons or mirror neutrons, may explain the matter-antimatter asymmetry observed in the Universe. Searches for this require an enormous and highly controlled flux of cold neutrons, which can be achieved at the European Spallation Source. Utilizing this, the HIBEAM and NNBAR program are therefore poised to open a new discovery window to neutron oscillation that surpasses previous discovery potentials by more than three orders of magnitude, and also opens up ancillary searches for e.g.\ axions.

\vspace{6pt} 

\textbf{Funding:} The author gratefully recognizes the contributions from the Olle Engkvists Stiftelse.

%\end{adjustwidth}

\end{document}